\documentstyle[twocolumn,aps,prl,epsf,floats]{revtex}
\begin{document}
\draft%
\preprint{Draft copy --- not for distribution}
%
%
\wideabs{
%
%
\title{Optical studies of charge dynamics in {\it c}-axis oriented superconducting
 \boldmath MgB$_2$ \unboldmath films}

%
%
\author{J.J.~Tu,$^{1,}$\cite{correspond} G.L.~Carr,$^2$ V.~Perebeinos,$^1$ C.C.~Homes,$^1$
M.~Strongin,$^1$ P.B.~Allen,$^3$ W.N.~Kang,$^4$ Eun-Mi~Choi,$^4$ Hyeong-Jin~Kim,$^4$
and Sung-Ik~Lee$^4$ }
%
%
\address{$^1$Department of Physics, Brookhaven National Laboratory, Upton,
New York 11973-5000}
\address{$^2$NSLS, Brookhaven National Laboratory, Upton, New York 11973-5000}
\address{$^3$Department of Physics and Astronomy, SUNY Stony Brook, Stony Brook
 NY 11794-3800}
\address{$^4$National Creative Research Initiative Center for Superconductivity,
 Department of Physics, Pohang University of Science and Technology, Pohang 790-784, Korea}

\date{\today}
\maketitle
%
%
\begin{abstract}
Temperature dependent optical conductivities and DC resistivity of {\it c}-axis
oriented superconducting (T$_c$ = 39.6~K) MgB$_2$ films ($\sim 450$~nm) have
been measured. The normal state {\it ab}-plane optical conductivities can be described
by the Drude model with a temperature independent Drude plasma frequency of 
$\omega_{p,D}=13\,600\,\pm 100$~cm$^{-1}$ or $1.68\pm 0.01$~eV. The normal state 
resistivity is fitted by the Bloch-Gr\"{u}neisen formula with an electron-phonon coupling 
constant $\lambda_{tr} = 0.13\pm 0.02$.  The optical conductivity spectra below T$_c$ of 
these films suggest that MgB$_2$ is a multi-gap superconductor.
\end{abstract}
%
%
}
%
%
\makeatletter%
\global\@specialpagefalse
\def\@oddhead{\hfil Preprint --- Submitted to PRL on July 2, 2001}
\let\@evenhead\@oddhead
\makeatother
\narrowtext%
\vspace*{-0.8cm}
%
%
The recent discovery of superconductivity in MgB$_2$ with T$_c$ of 39~K has generated 
much scientific interest \cite{nagamatsu01}.  As in the case of the high-T$_c$ cuprates, 
debate rages as to the mechanism of superconductivity in this material.  Initial isotope 
effect measurements suggested electron-phonon coupling as the pairing mechanism for
superconductivity in MgB$_2$ \cite{budko01,hinks01}.  Many theoretical studies
\cite{kortus01,amy01,kong01,an01} since then have concluded that strong
electron-phonon coupling is responsible for the high transition temperature, with 
$\lambda\sim 1$.  However, other pairing mechanisms have also been proposed, e.g. 
`dressing' and `undressing' of holes \cite{hirsch01}, acoustic plasmons \cite{voelker01} and 
the `filamentary' thoery \cite{phillips01}.  This inconclusive state of affairs is mainly 
due to the lack of consensus on many important physical quantities in MgB$_2$.  For example, 
the reported values for the superconducting gap $2\Delta$ vary from 4~meV \cite{bollinger01} 
to 15 meV \cite{giubileo01}. Infrared spectroscopy is able to measure such quantities as
the scattering rate $1/\tau$, the Drude plasma frequency $\omega_{p,D}$ and $2\Delta$ 
\cite{tinkham}. In this work, we analyze the optical data of MgB$_2$ to determine 
the electron-phonon coupling constant, $\lambda_{tr}$, in a similar fashion as in the 
optical study \cite{puchkov94} of Ba$_{0.6}$K$_{0.4}$BiO$_3$ (T$_c\sim 30$~K), where 
$\lambda_{tr} \sim 0.2$ was obtained experimentally.

There have been very few optical studies on MgB$_2$ to date.  Gorshonov {\it et
al.} \cite{gorshunov01} measured the reflectance of a polycrystalline pellet
using the grazing angle method and set a lower limit of $2\Delta$ to be $3 - 4$
meV.  Pronin {\it et al.} \cite{pronin01} examined the complex optical
conductivity of a MgB$_2$ thin film in the frequency range of $0.5 - 4$~meV.  
More recently, Jung {\it et al.} \cite{jung01} carried out transmission measurements 
on a {\it c}-axis oriented MgB$_2$ film ($\sim 50$~nm) with T$_c\sim 33$~K and fitted 
the data with a gap value of $2\Delta(0)\sim 5.2$~meV.  However, to obtain the optical constants 
of bulk MgB$_2$ in a wide frequency region, reflectivity measurements are the preferred method.

In this Letter, temperature dependent optical conductivities and DC resistivity
of {\it c}-axis oriented superconducting (T$_c= 39.6$~K) MgB$_2$ films ($\sim
450$~nm) are reported.  The normal state {\it ab}-plane optical conductivities
can be well described by the Drude model with $\omega_{p,D}=13600\pm 100$~cm$^{-1}$.  
Using this plasma frequency $\lambda_{tr}=0.13\pm 0.02$ is determined by fitting the 
DC resistivity data. In addition, the optical conductivities in the superconducting state 
exhibit complex behavior suggesting that MgB$_2$ is a multi-gap superconductor.

For this study, several {\it c}-axis oriented MgB$_2$ films are used: one
very thin film ($\sim 50$~nm) similar to the film studied by Jung {\it et al.}
\cite{jung01} and two thicker films ($\sim 450$~nm). These high-quality {\it
c}-axis oriented films were deposited on {\it c}-cut Al$_2$O$_3$ substrates
using a pulsed laser deposition method as described previously \cite{kang01}.
X-ray measurements showed that the MgB$_2$ grains were highly oriented
with their {\it c}-axes normal to the substrate. These MgB$_2$ films have a tan
appearance, similar to the high purity MgB$_2$ polycrystalline samples \cite{budko01}.  
The thick MgB$_2$ films ($\sim 450$~nm) are opaque in the visible region.

The Al$_2$O$_3$ substrates with the MgB$_2$ films are mounted on an
optically-black cone, and the temperature dependent reflectance is measured in
a near-normal-incidence arrangement from $\sim 30$ to over $22\,000$ cm$^{-1}$,
with the electric field parallel to the {\it ab}-plane on Bruker IFS 66v/S
and 113v spectrometers.  The absolute reflectance is determined by evaporating a
gold film {\it in situ} in ultra-high vacuum ($\sim 10^{-8}$~Torr).  The details of 
this technique have been described previously \cite{homes93}.  The optical conductivities 
are then determined from a Kramers-Kronig analysis of the reflectance.

In Fig.~1(a), the temperature dependent DC sheet resistance, R$_\Box$, measured
by a standard four-probe technique of a MgB$_2$ film ($\sim 450$~nm) is shown.
The residual resistance ratio (RRR) of R$_\Box$ at 295~K and at 40~K is 2.2.  The
low temperature region near T$_c$ is given in the insert of Fig.~1(a).  The
superconducting transition in this film is extremely sharp with
a transition region of $\delta{\rm T}_c <0.1$~K and a T$_c$ of 39.6~K
indicating that these thick MgB$_2$ films are of excellent quality \cite{kang01}.

%
%
\begin{figure}[t]
\epsfxsize=5.75cm%
\vspace*{-0.2cm}%
\centerline{\epsffile{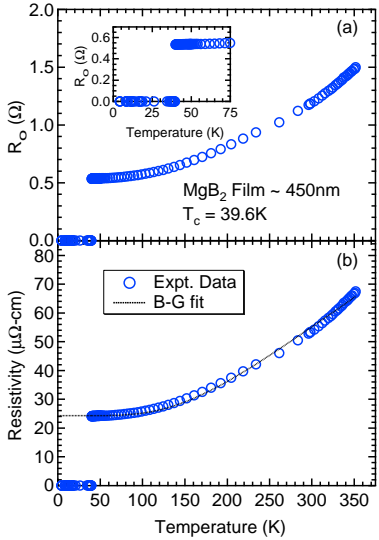}}%
\vspace*{-0.2cm}%
\medskip
\caption{The DC resistivity data of a {\it c}-axis
oriented MgB$_2$ film ($t\sim 450$~nm).  (a) Temperature dependent R$_\Box$
(open circles). Inset: low temperature region of R$_\Box$ near T$_c$; (b)
Temperature dependence of the resistivity together with a fit to the
Bloch-Gr\"{u}neisen formula. }
\end{figure}

The raw data of the optical measurements on these MgB$_2$ films ($\sim 450$~nm)
are summarized in Fig.~2.  The absolute reflectance is quite high as shown in Fig.~2(a), 
however, several sharp phonon features can be clearly identified.  As a comparison, the
reflectance of the thin MgB$_2$ film ($\sim 50$~nm) is also measured. The two strong
infrared active TO-phonons of {\it c}-cut Al$_2$O$_3$ crystals at 440 and 570~cm$^{-1}$
 \cite{barker63} can be easily observed for the thin MgB$_2$ film
but completely absent in the reflectance data of the thick films ($\sim 450$~nm),
indicating that the thick films are totally opaque. Therefore, the {\it ab}-plane optical
properties measured for these MgB$_2$ films ($\sim 450$~nm) are intrinsic.  In the insert,
the reflectance data at 295~K is given for the entire frequency region: from
30 to $22\,000$~cm$^{-1}$. The largest possible frequency interval is needed to
carry out a reliable Kramers-Kronig analysis.  The results of such an analysis are
shown as temperature dependent $\sigma_1(\omega)$ in Fig.~2(b) and
$\sigma_2(\omega)$ in Fig.~2(c).  Superconducting behavior can be easily
identified as a drop in $\sigma_1(\omega)$ at low frequencies below T$_c$.

%
%
\begin{figure}[t]
\epsfxsize=6.25cm%
\vspace*{-0.2cm}%
\centerline{\epsffile{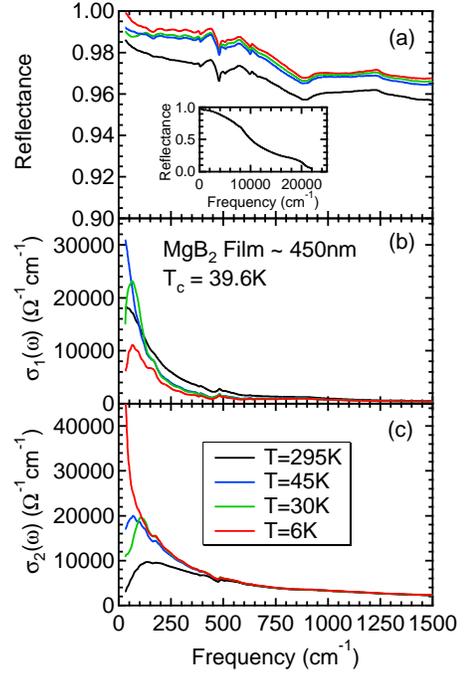}}%
\vspace*{-0.2cm}%
\medskip
\caption{The temperature dependent {\it ab}-plane optical data of {\it c}-axis
oriented MgB$_2$ films ($t\sim 450$~nm) from 30 to 1500 cm$^{-1}$.  (a) The
reflectance data showing sharp phonon modes. Inset: optical reflectance
spectrum for the entire frequency region at 295~K; (b) Temperature dependent
$\sigma_1(\omega)$; (c) Temperature dependent $\sigma_2(\omega)$. }
\end{figure}

The normal state optical conductivities of these MgB$_2$ films are analyzed in
Fig.~3.  In Fig.~3(a), $\sigma_1(\omega)$ and $\sigma_2(\omega)$ at 295~K are
shown.  Both the real and imaginary parts of the optical conductivities at low
frequencies can be well described by the simple Drude model of the form:
\begin{equation}
  \tilde\sigma(\omega)=\sigma_1+i\sigma_2=
  {1\over{4\pi}} { {\omega_{p,D}^2\tau} \over {1-i\omega\tau}}, \ \ \
  \omega_{p,D}^2={ {4\pi ne^2} \over {m^\ast} }
\end{equation}
where $\omega_{p,D}$ is the Drude plasma frequency, $1/\tau$ is the scattering
rate, $n$ is the number of free-carriers per unit volume and $m^\ast$ is the average 
effective mass of the occupied carrier states.  The Drude model
describes the experimental data surprisingly well at 295~K.  The fitting
parameters have the values $\omega_{p,D} = 13\,600 \pm 100$~cm$^{-1}$, and
$1/\tau = 170\pm 5$~cm$^{-1}$. This Drude plasma frequency of $13\,600$
cm$^{-1}$ is quite consistent with the value obtained from an optical study of
a polycrystalline MgB$_2$ sample \cite{kuzmenko01}. However, in addition to the
Drude peak, some other contributions to $\sigma_1(\omega)$  are also observed
in that optical study \cite{kuzmenko01}. Using the optical data, one can
determine the DC resistivity $\rho=1/\sigma_0$ to be $53\pm2$~$\mu\Omega$-cm at 295~K.
Thus, the averaged thickness of this MgB$_2$ film is derived as $t = \rho/$R$_\Box$ = $450\pm
20$~nm which agrees very well with the typical thickness of 400~nm of these films 
\cite{kang01}. The DC resistivity can now be plotted as shown in Fig.~1(b). It is 
interesting that the experimental Drude plasma frequency of 1.68~eV 
is much smaller than the value of $\sim 7$~eV predicted by calculations of the
electronic structure in MgB$_2$ \cite{kortus01,amy01,kong01,an01}. These calculations
usually give values of Drude plasma frequencies that are reasonably close
to experimental values, even in highly correlated systems like high T$_c$ cuprates
\cite{pickett88}.

Keeping $\omega_{p,D}$ the same, the optical conductivities at 45~K are fitted
with Eq.~(1). The results are given in Fig.~3(b).  Both $\sigma_1(\omega)$ and
$\sigma_2(\omega)$ again fit well with the Drude model with a scattering
rate of $1/\tau = 75\pm 5$~cm$^{-1}$. In addition, the DC resistivity at 45~K is in
good agreement with the zero frequency extrapolation of $\sigma_1(\omega)$.
Therefore, the DC conductivity and real part the optical conductivity are in 
excellent agreement.  

%
%
\begin{figure}[t]
\epsfxsize=6.25cm%
\vspace*{-0.2cm}%
\centerline{\epsffile{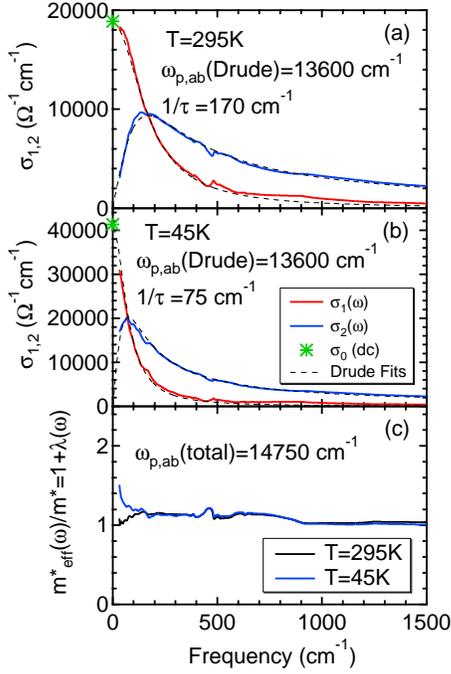}}%
\vspace*{-0.2cm}%
\medskip
\caption{The analysis of normal state {\it ab}-plane optical constants of
the {\it c}-axis oriented MgB$_2$ films. (a) Frequency dependent optical
conductivities: $\sigma_1(\omega)$ and $\sigma_2(\omega)$ at 295~K with the Drude fits; 
(b) Frequency dependent optical conductivities at 45~K; (c) Frequency dependent effective 
mass ratio for T$>T_c$.  }
\end{figure}

From the {\it ab}-plane optical data, one can calculate the frequency dependent
electron-phonon coupling constant $\lambda(\omega)$ in the extended Drude
formalism \cite{puchkov96}:
\begin{equation}
  {{m_{eff}^\ast(\omega)} \over {m^\ast}} = 1+\lambda(\omega) =
  {1\over{4\pi}} {{\omega_p^2}\over{\omega}} {\rm Im} \left[
  {1\over{\tilde\sigma(\omega)} } \right],
\end{equation}
where $\omega_p$ is the total plasma frequency of free {\it ab}-plane carriers.  
The purpose of casting the optical data in the extended Drude form
is to account for the small deviations from the simple Drude model by using a
frequency dependent scattering rate $1/\tau(\omega)$.  The result of this
analysis is shown in Fig.~3(c).  The value of $\lambda(\omega)$ derived
optically varies from 0 to about 0.2 in the optical phonon region, where
$\omega_p =14\,750\pm 150$ cm$^{-1}$ is derived from the conductivity sum rule.
The value of $\omega_p$ is slightly larger than $\omega_{p,D}$ due to the fact that 
the sum rule captures addtional spectral weight in the high frequency region.

The electron-phonon coupling constant $\lambda_{tr}$ is traditionally
determined from the temperature dependent DC resistivity using the
Bloch-Gr\"{u}neisen formula
\begin{equation}
  \rho(T)=\rho_0+\lambda_{tr} {{4\pi}\over{\omega_{p,D}^2}} {{128\pi(k_BT)^5}\over
  {(k_B\Theta_D)^4}}
  \int_0^{\Theta_D\over{2T}} {{x^5}\over{\sinh^2x}} dx,
\end{equation}
with three parameters: $\rho_0$ - the residual resistivity at $T = 0$;
$\Theta_D$ - the Debye temperature and $\lambda_{tr}$.  A non-linear least
squares fit to the resistivity data with the Eq.~(3) is given
in Fig.~1(b) using $\omega_{p,D} = 1.68\pm 0.01$~eV. The experimental curve and the 
theoretical fit agree quite well with the fitting parameters: $\rho_0 = 24.3\pm 
0.3$~$\mu\Omega$-cm; $\Theta_D = 950\pm 100$~K and $\lambda_{tr}=0.13\pm 0.02$.  
The value $\Theta_D = 950\pm 100$~K is consistent with the experimentally measured value 
that varies from 800~K \cite{kremer01} to 1050~K \cite{bouquet01}.  However, $\lambda_{tr}
=0.13\pm 0.02$ is significantly smaller than most theoretical predictions of
$\lambda\sim 1$ \cite{kortus01,amy01,kong01,an01} in MgB$_2$.

%
%
\begin{figure}[t]
\epsfxsize=6.25cm%
\vspace*{-0.2cm}%
\centerline{\epsffile{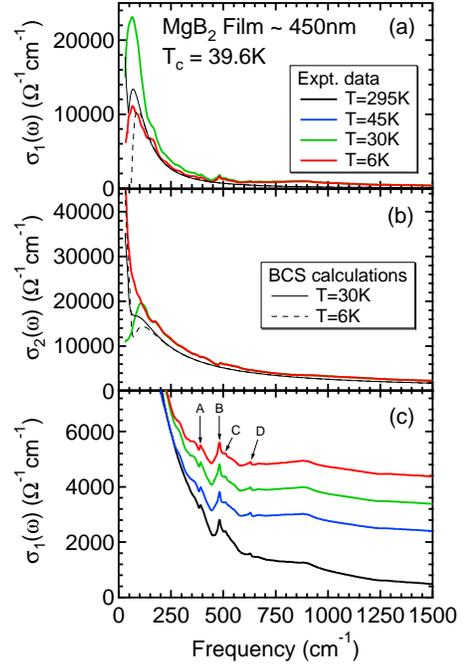}}%
\medskip
\vspace*{-0.2cm}%
\caption{The analysis of superconducting state {\it ab}-plane optical constants of the 
{\it c}-axis  MgB$_2$ films. (a) Temperature dependent $\sigma_1(\omega)$ (thick lines) with 
the BCS fits; (b) Temperature dependent $\sigma_2(\omega)$ with the BCS fits; (c) Temperature 
dependent $\sigma_1(\omega)$ showing four $\Gamma$-point phonons. The spectra corresponding to 
different temperatures are offset for clarity.   }
\end{figure}

The optical conductivities of these MgB$_2$ films in the superconducting state
are examined in Fig.~4.  The superfluid plasma frequency is found to be $\omega_{p,S} 
= 7300\pm 50$ cm$^{-1}$ at 6~K from the Ferrel-Glover-Tinkham sum rule.  However, the optical 
spectra below T$_c$ cannot be fitted by the BCS model using a single isotropic gap.  Theoretical
curves at 30~K and 6~K generated with a BCS model \cite{zimmermann91} are shown in Figs.~4(a) 
and 4(b).  The parameters used are: $2\Delta = 65$ cm$^{-1}$ and $1/\tau = 75$ cm$^{-1}$.  
There are significant deviations between the experimental data and the BCS calculations. 
However, from our optical data the upper and lower limits of the superconducting gap can be
estimated: 5~meV $< 2\Delta_x < 15$~meV.  The complex gap behavior observed 
in our optical conductivity data in the superconducting state adds support to the 
suggestion that MgB2 is a multi-gap superconductor \cite{amy01,giubileo01,bouquet01}.

Four sharp phonon peaks can be identified in $\sigma_1(\omega)$ as shown in
Fig.~4(c) that can be assigned to $\Gamma$-point optical phonons in MgB$_2$
\cite{kortus01}. The two strong phonon peaks marked as A and B are the two infrared active 
lattice modes: at 380 cm$^{-1}$ ($E_{1u}$) and at 480 cm$^{-1}$ ($A_{2u}$). Their relatively 
large oscillator strengths are the consequence of the low plasma frequency in
MgB$_2$. Two weak phonon peaks marked as C and D at 510 cm$^{-1}$ and 630
cm$^{-1}$ are tentatively assigned as the Raman active $E_{2g}$ mode and the
silent $B_{1g}$ mode according to the phonon calculations \cite{kortus01}. These two phonons 
with even symmetry become infrared active because of the lattice imperfections in the films. 
Alternatively, several Raman studies \cite{chen01} on MgB$_2$ have assigned a very broad band 
centered at 620 cm$^{-1}$ as the $E_{2g}$ mode. In addition, three broad features are also
observed at 160, 880 and 1240 cm$^{-1}$ in $\sigma_1(\omega)$. The resolution of
this optical study is 4~cm$^{-1}$ in the phonon region, and none of the four
sharp phonon modes exhibit detectable changes in either their intensities, peak
positions, or line-widths going through T$_c$.

The surprising aspect of our results is the small value of $\lambda_{tr}=0.13$
derived from both the DC resistivity and optical conductivity measurements.  A
simple application of McMillan formula \cite{mcmillan68} with $\lambda \cong
\lambda_{tr}=0.13$ will give T$_c < 1$~K.  However, there are several reasons
why the BCS theory should not be abandoned right way for MgB$_2$: 1) the superconducting
gap in MgB$_2$ has unusual properties.  Gap anisotropy including dimensional effects 
\cite{allen82} modifies T$_c$ relative to the McMillan formula; 2) the $\lambda$ value that 
goes into the McMillan formula can differ somewhat with respect to $\lambda_{tr}$ \cite{allen00}; 
3) c-axis optical and transport properties should be experimentally studied. On the other hand, 
given the small value of $\lambda_{tr} =0.13$ alternative mechanisms of superconductivity in 
MgB$_2$ should be examined both experimentally and theoretically.  It is interesting to note 
that many of the optical constants in MgB$_2$ are quite similar to those in 
Ba$_{0.6}$K$_{0.4}$BiO$_{3}$ \cite{puchkov94}, e.g. the scattering rate, the Drude plasma 
frequency, and particularly the small value of $\lambda_{tr}$.  A common mechanism might be
responsible for superconductivity in both systems. In addition, having a small free-carrier 
plasma frequency ($< 3$~eV) seems to be an universal characteristic shared by almost all 
superconductors with a T$_c > 30$~K.

In conclusion, we have measured optical conductivities and DC resistivity of
c-axis oriented superconducting MgB$_2$ films.  With a Drude plasma frequency of 
$\omega_{p,D} = 13\,600\pm 100$~cm$^{-1}$, $\lambda_{tr}=0.13\pm 0.02$ is determined 
from DC resistivity data.  The small measured $\lambda_{tr}$ value poses a serious problem 
to the strong electron-phonon coupling picture. Other theoretical models need to be explored
to account both for the complex behavior of the superconducting gap and possible different
pairing mechanism in MgB$_2$.

%
%
We thank P.C. Canfield, V.J. Emery, J.E.Hirsch, P.D. Johnson,  S.A. Kivelson, G. Schneider, 
T. Valla, T. Vogt, and Z. Yusof for helpful discussions.  Part of the work was supported 
by the U.S. Department of Energy under Contract No. DE-AC02-98CH10886 and the other part by 
the Ministry of Science and Technology of Korea through the Creative Research Initiative Program.  
Research undertaken at NSLS was supported by the U.S. DOE, Division of Materials and Chemical 
Sciences.
\vspace*{-0.25cm}
%
%

%
%
%

\begin{references}
%
\vspace*{-1.0cm}
%
\bibitem[*]{correspond} Electronic address: jtu@bnl.gov

\bibitem{nagamatsu01} J.~Nagamatsu {\it et al.}, Nature (London) {\bf 410}, 63 (2001).

\bibitem{budko01} S.L.~Bud'ko {\it et al.}, \prl {\bf 86}, 1877 (2001).

\bibitem{hinks01} D.G.~Hinks, H.~Claus, and J.D.~Jorgensen, cond-mat/0104242.

\bibitem{kortus01} J.~Kortus {\it et al.}, \prl {\bf 86}, 4656 (2001).

\bibitem{amy01}  A.Y.~Liu, I.I.~Mazin, and J.~Kortus, cond-mat/0103570.

\bibitem{kong01} Y.~Kong, O.V. Dolgov, O. Jepsen, and O.K.~Anderson, \prb
  {\bf 64}, 020501(R) (2001).

\bibitem{an01} J.M.~An and W.E.~Pickett, \prl {\bf 86}, 4366 (2001).

\bibitem{hirsch01} J.E.~Hirsh and F.~Marsiglio, cond-mat/0102479.

\bibitem{voelker01} K.~Voelker, V.I.~Anisimov, and T.M.~Rice, cond-mat/0103082.

\bibitem{phillips01} J.C.~Phillips and J.~Jung, cond-mat/0102261.

\bibitem{bollinger01} G.~Rubio-Bollinger, H.~Suderow, and S.~Vieira, \prl
  {\bf 86}, 5582 (2001).

\bibitem{giubileo01} F.~Giubileo {\it et al.}, cond-mat/01045146.

\bibitem{tinkham} M.~Tinkham, {\it Introduction to Superconductivity}
  (Krieger, Malabar, 1975); B.~Farnworth and T.~Timusk, \prb {\bf 10}, 5119
  (1976); F.~Gao {\it et al.}, \prb {\bf 54}, 700 (1996).

\bibitem{puchkov94} A.V.~Puchkov, T.~Timusk, W.D.~Mosley, and R.N.~Shelton, \prb
  {\bf 50}, 4144 (1994).

\bibitem{gorshunov01} B.~Gorshunov {\it et al.}, cond-mat/0103164.

\bibitem{pronin01} A.V.~Pronin, A.~Pimenov, A.~Loidl, and S.I.~Kransnosvobodtsev,
cond-mat/0104291.

\bibitem{jung01} J.H.~Jung {\it et al.}, cond-mat/0105180.

\bibitem{kang01} W.N.~Kang {\it et al.}, Science {\bf 292}, 1521 (2001).

\bibitem{homes93} C.C.~Homes, M.~Reedyk, D.~Crandles, and T.~Timusk,
  Appl. Opt. {\bf 32}, 2972 (1993).

\bibitem{barker63} A.S.~Barker, Phys. Rev. {\bf 132}, 1474 (1963).

\bibitem{kuzmenko01} A. B. Kuz'menko {\it et al.}, in {\it Sixth International
 Conference on Spectroscopy of Novel Superconductors}, edited by A. Bansil
 (Elsevier, Chicago, 2001), p.~51 (poster P28).

\bibitem{pickett88} W.E.~Pickett, P.B.~Allen, and H.~Krakauer, \prb {\bf 37},
 7482 (1988).

\bibitem{puchkov96} A.V.~Puchkov, D.N.~Basov, and T.~Timusk, J.~Phys.~C {\bf 8},
   $10\,049$ (1996).

\bibitem{kremer01} R.K.~Kremer, B.J.~Gibson, and K.~Ahn, cond-mat/0102432.

\bibitem{bouquet01} F.~Bouquet {\it et al.}, cond-mat/0104206.

\bibitem{zimmermann91} W.~Zimmermann {\it et al.}, Physica C {\bf 183}, 99 (1991).

\bibitem{chen01} X. K. Chen {\it et al.}, cond-mat/0104005; A.F.~Goncharov {\it
 et al.}, cond-mat/0104042; J.~Hlinka {\it et al.}, cond-mat/0105275.

\bibitem{mcmillan68} W.L.~McMillan, Phys. Rev. {\bf 167}, 331 (1968);
 P.B.~Allen and R.C.~Dynes, Phys. Rev. B {\bf 12}, 905 (1975).

\bibitem{allen82} P.B.~Allen, Z.~Phys.~B {\bf 47}, 45 (1982) (and references therein).

\bibitem{allen00} P.B.~Allen, in {\it Handbook of Superconductivity}, edited by
 C.P.~Poole (Academic, San Diego, 2000), p.~478.

%
\end{references}
\end{document}